\documentclass{PoS}

\usepackage{epstopdf}
\usepackage{amsmath,amssymb}
\usepackage{tikz}

\newcommand{\be}{\begin{equation}}
\newcommand{\ee}{\end{equation}}
\newcommand{\ba}{\begin{eqnarray}}
\newcommand{\ea}{\end{eqnarray}}

\def\lsim{\raise0.3ex\hbox{$<$\kern-0.75em\raise-1.1ex\hbox{$\sim$}}}
\def\gsim{\raise0.3ex\hbox{$>$\kern-0.75em\raise-1.1ex\hbox{$\sim$}}}

\title{
{\vspace{-15mm} \normalsize\hfill{\small CERN-PH-TH/2013-257}}\\[20mm]
The chiral phase transition for two-flavour QCD at imaginary and zero chemical potential}

\ShortTitle{$N_f=2$ chiral transition}

\author{Claudio Bonati, Massimo D'Elia \\
Dipartimento di Fisica, Universit\`a di Pisa and INFN, 
Sezione di Pisa, \\ Largo Pontecorvo 3, 56127 Pisa, Italy \\
E-mail: \email{bonati@df.unipi.it}}

\author{Philippe de Forcrand\\
Institute for Theoretical Physics, ETH Z\"urich, CH-8093 Z\"urich, Switzerland and\\
CERN, Physics Department, TH Unit, CH-1211 Geneva 23, Switzerland \\ 
E-mail: \email{forcrand@phys.ethz.ch}} 

\author{\speaker{Owe Philipsen} \\
        Institut f\"ur Theoretische Physik, Goethe-Universit\"at Frankfurt, \\ 60438
Frankfurt am Main, Germany \\
        E-mail: \email{philipsen@th.physik.uni-frankfurt.de}}
        
\author{Francesco Sanfilippo \\
Laboratoire de Physique Th\'eorique (B\^at.~210)
Universit\'e  Paris Sud, Centre d'Orsay,\\ F-91405
Orsay-Cedex, France, and \\
INFN Sezione di Roma, P.le Aldo Moro 5, 00185 Roma, Italy \\
E-mail: \email{francesco.sanfilippo@th.u-psud.fr}}

\abstract{
The chiral symmetry of QCD with two massless quark flavours gets restored in a non-analytic chiral phase transition at finite temperature and zero density. Whether this is a first-order or a second-order transition  has not yet been determined unambiguously, due to the difficulties of simulating light quarks. We investigate the nature of the chiral transition as a function of quark mass and imaginary chemical potential, using staggered fermions on $N_t=4$ lattices. At sufficiently large imaginary chemical potential, a clear signal for a first-order transition is obtained for small masses, which weakens with decreasing imaginary chemical potential. The second-order critical line $m_c(\mu_i)$, which marks the boundary between first-order and crossover behaviour, extrapolates to a finite $m_c(\mu_i=0)$ with known critical exponents. This implies a definitely first-order transition in the chiral limit on relatively coarse, $N_t=4$ lattices.
}

\FullConference{31st International Symposium on Lattice Field Theory - LATTICE 2013\\
		July 29 - August 3, 2013\\
		Mainz, Germany}

\begin{document}

\section{Introduction}

Because of its relevance to heavy ion collisions, the early universe and astrophysics, the order of the QCD
phase transition as a function of temperature and baryon chemical potential is a subject of intense numerical 
studies. In order to understand the interplay of the centre and chiral symmetries and their respective breaking it
is necessary to also study the phase transition as a function of
the number of quark flavours and the quark masses. Because of the well-known sign problem,
reliable information on the phase diagram at finite density remains very sketchy to date \cite{revPF,revOP}.
Thus the dependence of the phase diagram on the microscopic or even unphysical parameters like imaginary chemical
potential, for which there is no sign problem, becomes highly relevant as a constraint on the 
physical phase diagram.

\begin{figure}[t]
\vspace*{-0.5cm}
\centerline{
\includegraphics[width=0.4\textwidth]{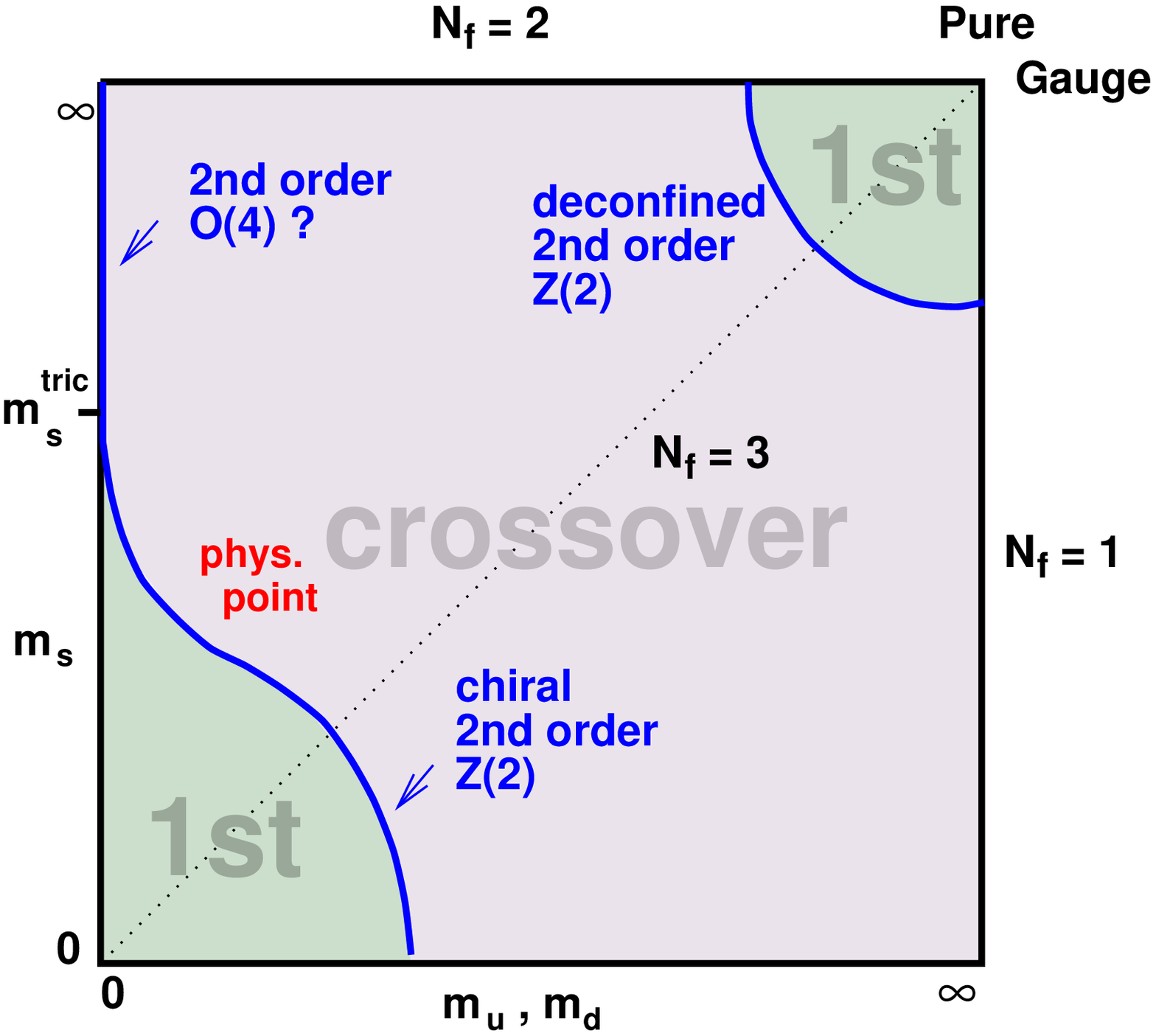}
\hspace*{0.5cm}
\includegraphics[width=0.6\textwidth]{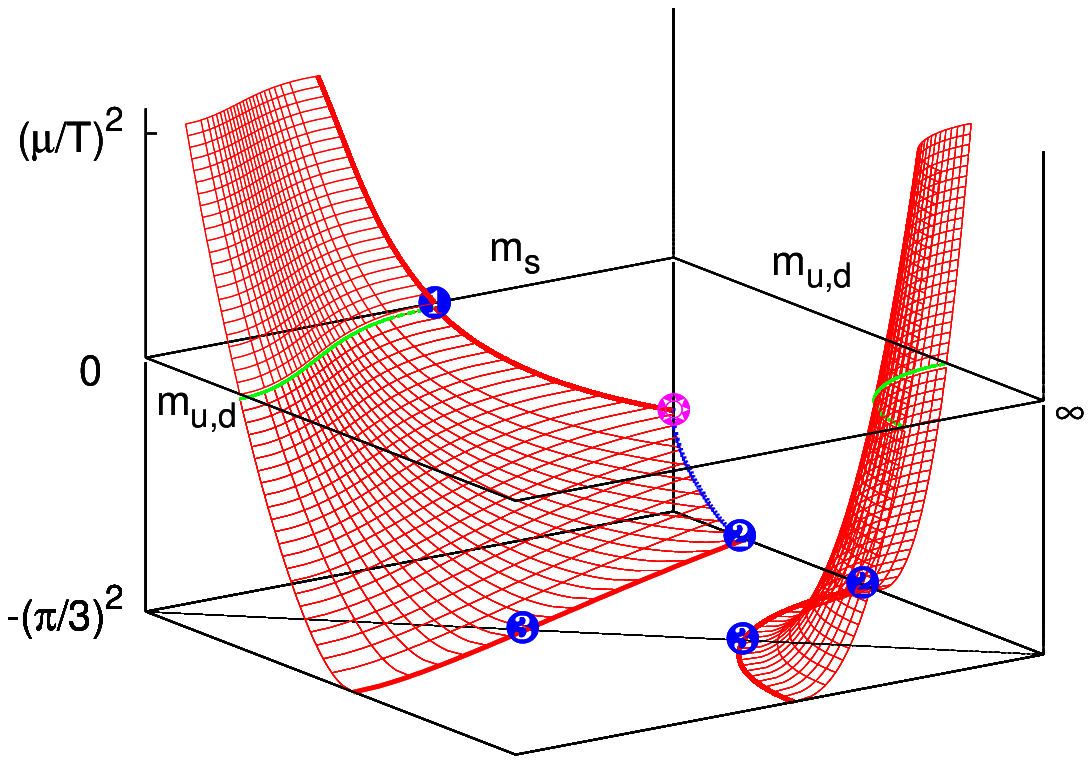}
}
\caption{({\em Left}) Schematic phase transition behaviour of $N_f=2+1$
QCD for different choices of quark masses $(m_{u,d},m_s)$ at
$\mu=0$. ({\em Right}) The same with chemical potential $\mu$ for quark number as an additional parameter. 
The critical boundary lines sweep out surfaces 
as $\mu$ is turned on.  At imaginary chemical potential $\mu=i\pi/3 T$, the critical surfaces terminate
in tricritical lines, which determines their curvature through critical scaling. 
}
\label{fig:schem}
\vspace*{-0.5cm}
\end{figure}

The current state of knowledge at zero density is summarised in Fig.~\ref{fig:schem} (left).
In the limits of zero and infinite quark mass for three degenerate flavours, QCD displays a first
order phase transition associated with the breaking of the chiral and centre symmetries, respectively.
For finite quark masses, these symmetries are broken explicitly and the associated first-order transitions
weaken away from these limits, until they disappear at critical points belonging to the $Z(2)$ 
universality class.
Continuum extrapolated results with improved staggered quarks tell us that the physical point is in the
crossover region \cite{nature}. The chiral \cite{chcrit} and deconfinement \cite{deccrit,saito} critical lines are known on coarse $N_t=4$ lattices, based on simulations with staggered and Wilson fermions, respectively.

At finite chemical potential, the critical lines sweep out surfaces, as shown in Fig.~\ref{fig:schem} (right). 
Both regions of chiral \cite{chcrit,mu4} and deconfinement \cite{deccrit,saito1} transitions shrink with real chemical potential and grow with
imaginary chemical potential
 
\begin{figure} 
%\centerline{
\vspace*{-0.5cm}
\begin{tikzpicture}
\draw (0,4) node [anchor=north] {$1^\text{st}$ order};
\draw[very thick] (0,4) -- (2.3,4);
\draw[very thick, dashed,->] (2.3,4) -- (5,4);
\filldraw (2.3,4) circle (2.5pt);
\draw (2.3,4) node [anchor=north] {$Z(2)$};
\draw (0,4.4) node [anchor=south] {$m_{\pi}=0$};
\draw (2.3,4) node [anchor=south] {$m_{\pi,c}$};
\draw (2,5.5) node [anchor=south] {$2^\text{nd}$ order $O(4)$ or $U(2)\times U(2)/U(2)$};
\draw[very thick, ->,dashed] (0,5.5) -- (5,5.5);
\filldraw (0,5.5) circle (2.5pt);
\draw (5,4.4) node [anchor=south] {$m_\pi$};
\end{tikzpicture}
\hspace*{0.5cm}
\includegraphics[width=0.5\textwidth]{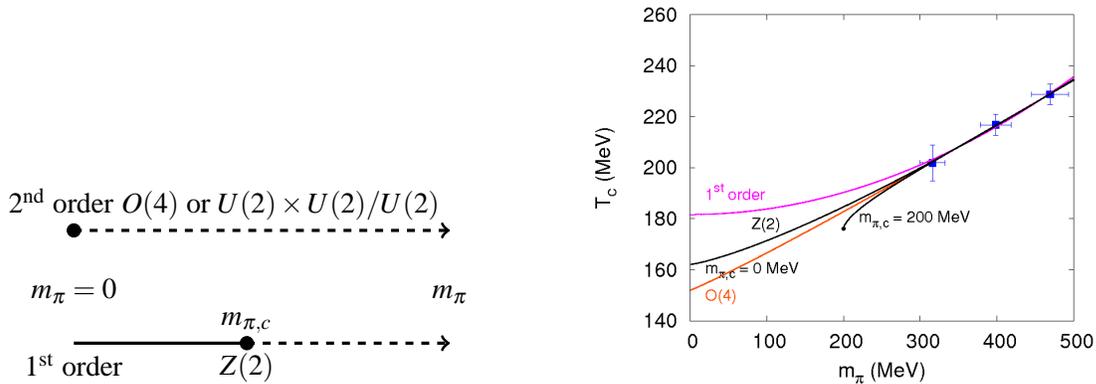}
%}
\caption[]{({\em Left}) Possible scenarios for the chiral phase transition as function of small pion mass.\\
({\em Right)} Chiral extrapolation of the pseudo-critical temperature for twisted mass Wilson fermions \cite{tmft}. 
}
\label{fig:nf2}
\vspace*{-0.5cm}
\end{figure}

However, the order of the thermal transition in the two flavour chiral limit (upper left corner of 
Fig.~\ref{fig:schem} (left)) remains open so far, with a long
history of conflicting lattice results between Wilson \cite{Tsukuba} and even within staggered \cite{DiG,Unger} fermions. 
The possibilities for continuum QCD have been discussed systematically in \cite{pw}. The classical action has a global $SU(2)_R\times SU(2)_L\times U_A(1)$ chiral symmetry, with the $U_A(1)$ being anomalous. 
Since the phase transition is associated with the breaking of a global symmetry, an analytic crossover
is ruled out, leaving room for either a first-order or second-order transition, the latter of 
$SU(2)_L\times SU(2)_R\simeq O(4)$ universality.
It has been pointed out that $U(2)_R\times U(2)_L/U(2)_V$ is in fact another possible universality class for a second-order transition \cite{vicari}. Once finite quark masses
are switched on, a second-order transition disappears immediately while a first-order transition gets
gradually weakened to disappear in a $Z(2)$ critical point. The two situations are sketched in Fig.~\ref{fig:nf2} (left). 
One of the deciding factors between the two scenarios is the strength of the axial anomaly, see
the discussions in \cite{pw,vicari,vicari2,mehta}.

Distinguishing between these two 
scenarios in lattice simulations is notoriously difficult since chiral fermions cannot be simulated.  
The standard strategy thus is to simulate at a number of finite quark masses and test for consistency 
with scaling relations which are valid when a second-order transition is approached. An example based on calculations
of the pseudo-critical temperature as a function of the pion mass for twisted mass Wilson fermions 
\cite{tmft} is shown in Fig.~\ref{fig:nf2} (right). In a neighbourhood of a second-order point, the
critical temperature is approached as 
\begin{equation}
T_c(m_\pi) = T_c(0) + A \cdot m_\pi^{2/(\beta\delta)}\;. \label{eq:chiralfit}
\end{equation}
As the figure illustrates, there is no way of distinguishing the different scenarios reliably in this manner.
The reason is that the critical exponents are very close, $1/(\beta\delta)=0.537,0.638$
for $O(4),Z(2)$, respectively. With current errorbars, even pion masses with half the physical value would
not be sufficient. The situation is similar for other scaling tests based on, e.g.~, the magnetic equation of
state or finite size scaling. 
We therefore propose to identify the chiral transition at imaginary chemical potentials, where it extends to finite quark masses, and utilise known scaling behaviour of the chiral critical surface in the 
$N_f=2$ plane of Fig.~\ref{fig:schem} (right) to arrive at a controlled chiral extrapolation, as explained in 
the following. Preliminary results have been presented in \cite{pos}.

\section{Phase diagram for imaginary chemical potential}

\begin{figure}[t]
\begin{center}
\centerline{
\includegraphics[width=0.318\textwidth]{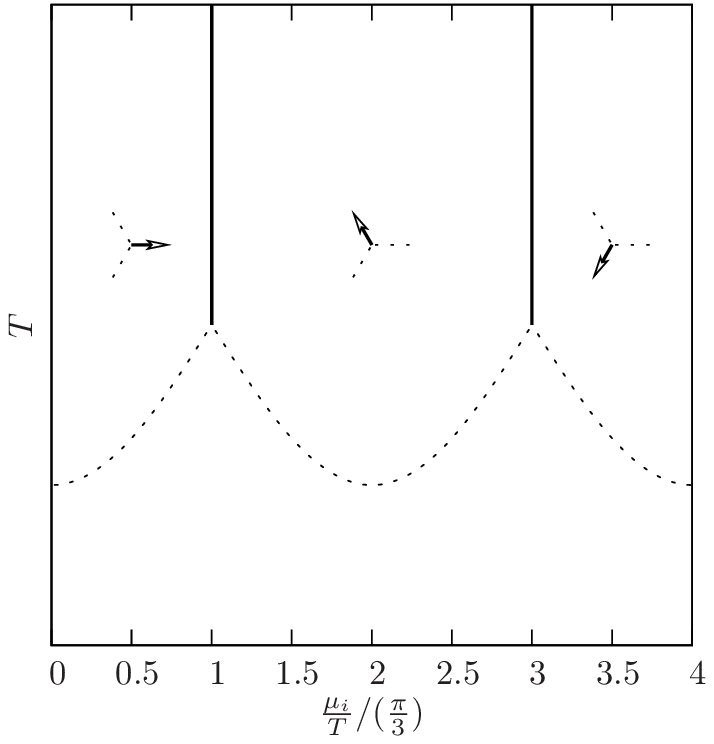}
\hspace*{0.5cm}
\includegraphics[width=0.32\textwidth]{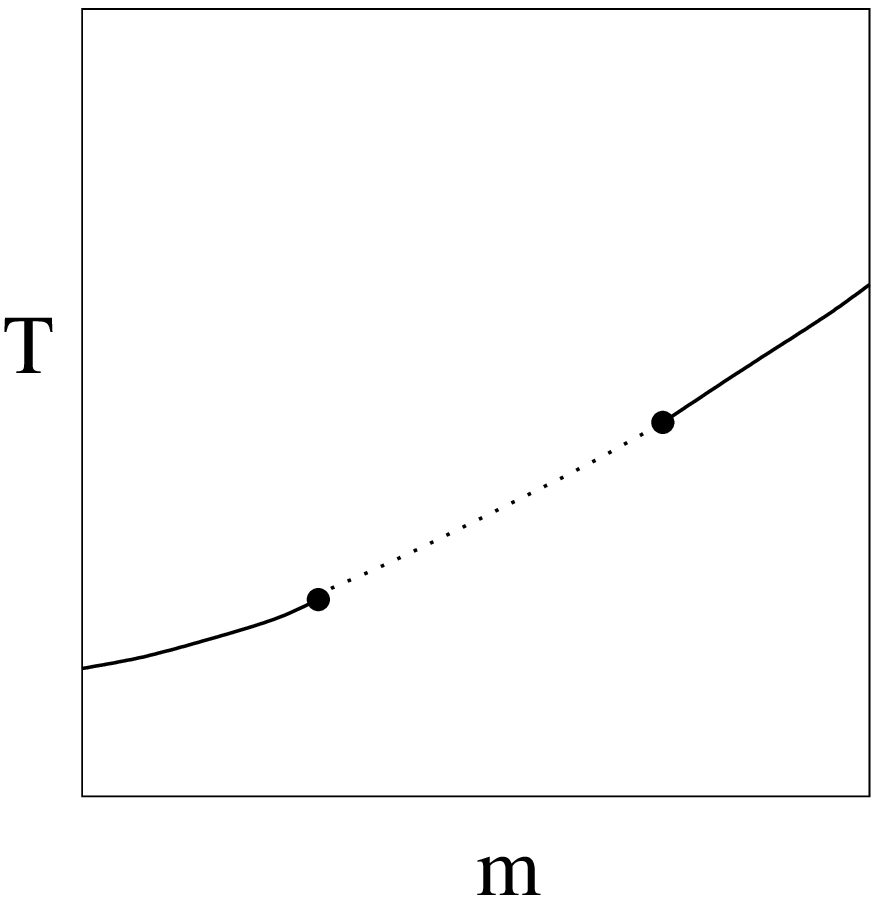}
\put(-64,83){\color{red}\small Tricritical}
\put(-97,35){\color{red}\small Tricritical}
\hspace*{0.5cm}
\includegraphics[width=0.32\textwidth]{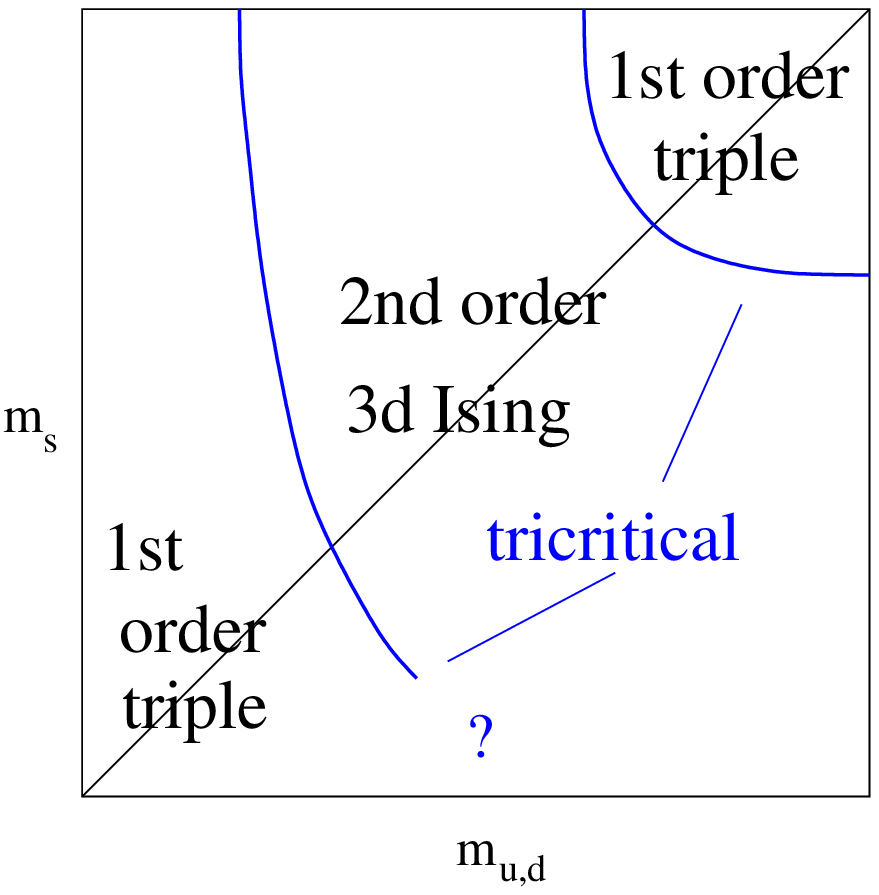}
\put(-49,136){\color{blue}$\bullet$}
\put(-102,136){\color{blue}$\bullet$}
\put(-38,102){\color{blue}$\bullet$}
\put(-88,51){\color{blue}$\bullet$}
}
\caption{({\em Left}) Generic phase diagram as a function of imaginary chemical potential and temperature. Solid lines are first-order Roberge-Weiss transitions. The behaviour along dotted lines depends on the number of flavours and the quark masses. ({\em Middle}) For $N_f=2$ and $N_f=3$, the endpoint of the Roberge-Weiss line is a triple point (where 3 first-order lines meet) for light or heavy quark masses,
and an Ising critical point for intermediate quark masses. Thus, two tricritical masses exist. 
({\em Right}) In the simplest scenario the calculated $N_f=2$ and $N_f=3$ tricritical points (bullets) 
are joined by
tricritical lines~\cite{OPRW}. 
\label{fig:rw}}
\end{center}
\vspace*{-0.5cm}
\end{figure}
There are two exact symmetries of the partition function,
\be
Z(\mu)=Z(-\mu), ~~~ Z\left(\frac{\mu}{T}\right) = Z\left(\frac{\mu}{T} + i\frac{2\pi n}{3}\right)\;,
\ee
which imply reflection symmetry in the imaginary $\mu$ direction about the ``Roberge-Weiss'' values 
$\mu = i\pi T/3 (2n+1)$ which separate different sectors of
the centre symmetry~\cite{RW}.
Transitions between neighbouring sectors
are of first-order for high $T$ and analytic crossovers
for low $T$ \cite{RW,fp1,el1},
as indicated Fig.~\ref{fig:rw} (left). 
The corresponding first-order transition lines may end
with a second-order critical point, or with a triple point, branching off into two first-order lines.
Which of these two possibilities occurs depends on the number of flavors and the quark masses. 

Recent numerical studies have shown that a triple point is found for heavy and light quark masses,
while for intermediate masses one finds a second-order endpoint. As a function of the quark mass, the phase diagram at $\mu/T = i\pi/3$ is as sketched Fig.~\ref{fig:rw} (middle).
This happens for both $N_f=2$ \cite{DEliaRW} and $N_f=3$ \cite{OPRW}. 
If one assumes that the $N_f=2$ and $N_f=3$ tricritical points are connected to each other in the
$(m_{u,d},m_s)$ quark mass plane, the resulting phase diagram is depicted Fig.~\ref{fig:rw} (right),
with two tricritical lines separating regions of first-order and of second-order transitions.
This phase diagram is the equivalent of Fig.~\ref{fig:schem} (left), now at imaginary chemical potential
$\mu/T = i\pi/3$. Note that the assumption of continuity of the tricritical lines can be checked
directly 
by numerical simulations, since there is no sign problem for imaginary $\mu$.

\section{Tricritical scaling}

For a given flavour configuration, the chiral critical surface is projected to a plane
and describes a critical line, whose lower end terminates in a tricritical point.
The important observation is that in the vicinity of a tricritical point scaling laws apply. In particular,
the functional form of a critical line extending from a tricritical point in the Roberge-Weiss plane upwards  is for the two-flavour theory 
\be
\left[(\mu/T)^2 + (\pi/3)^2\right] \propto (m_{u,d}-m_{\rm tric})^{5/2}\;.
\label{eq1}
\ee
Note that the scaling exponents governing the behaviour near the tricritical point are mean-field,
because QCD becomes 3-dimensional as the correlation length diverges while the temperature is fixed
to that of the tricritical point, and $d=3$ is the upper critical dimension for tricriticality.

\begin{figure}[t]
\begin{center}
\centerline{
\includegraphics[width=0.55\textwidth]{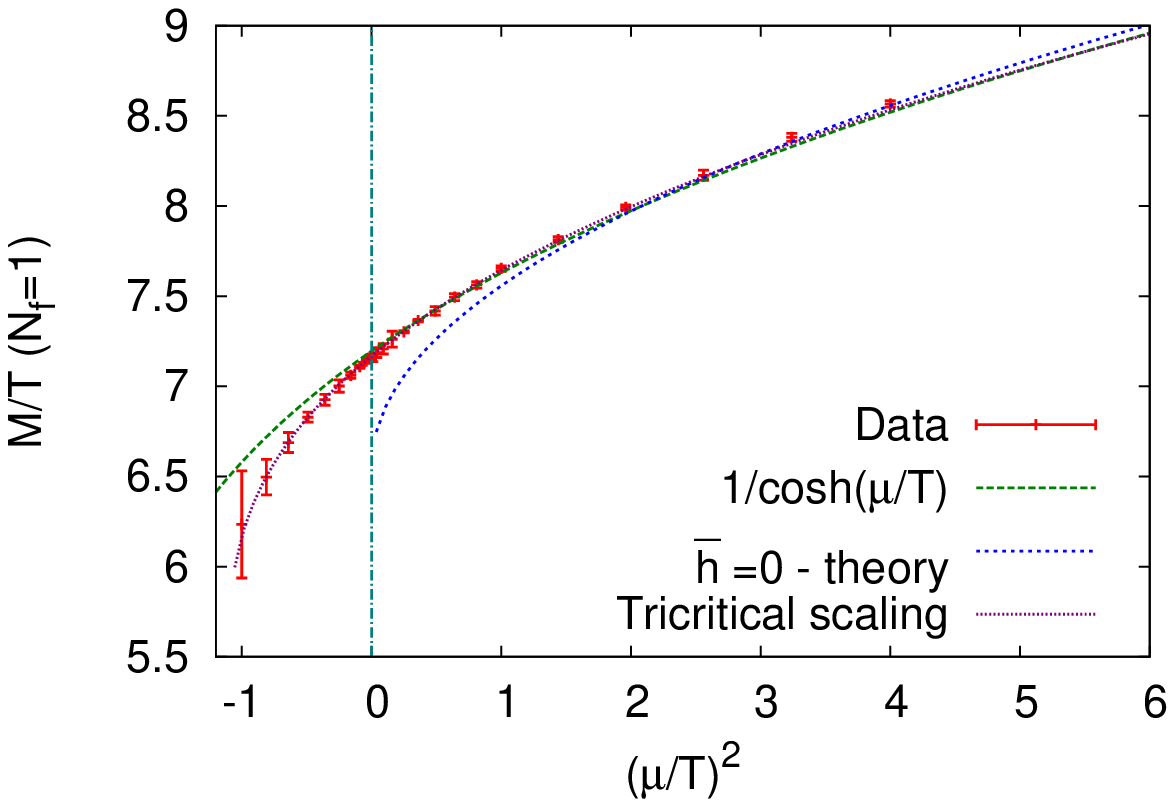}
\hspace*{0.2cm}
\includegraphics[width=0.55\textwidth]{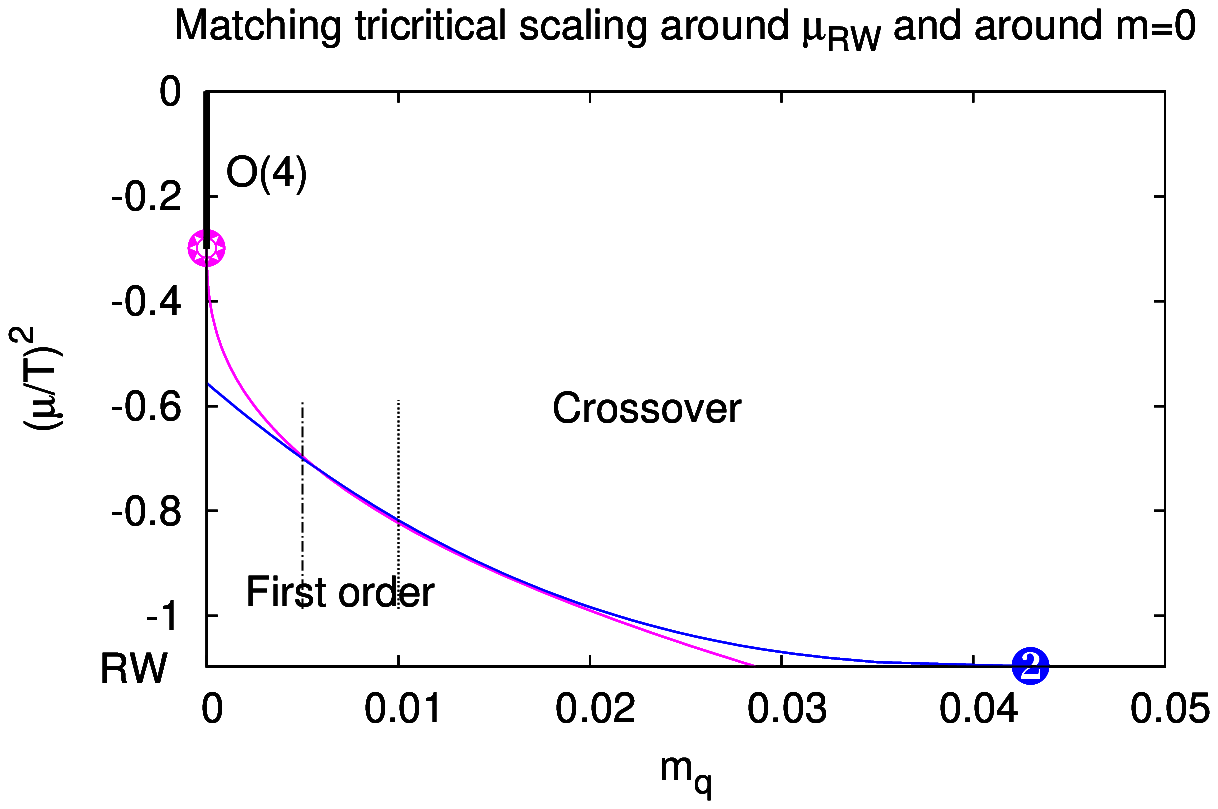}
}
\caption[]{({\em Left}) For heavy quarks, tricritical scaling of the deconfinement critical line in the vicinity of the
Roberge-Weiss imaginary-$\mu$ value extends far into the region of real $\mu$~\cite{deccrit}.
({\em Right}) The $N_f=2$ backplane of Fig.~\ref{fig:schem} (right). Two tricritical points at $m=0$ and 
$\mu=i\pi T/3$ are connected by a $Z(2)$ critical line, which obeys tricritical scaling in the vicinity
of the tricritical points.
\label{fig:mc}}
\end{center}
\vspace*{-0.5cm}
\end{figure}
Fig.~\ref{fig:mc} (left) shows the deconfinement critical mass as a function of imaginary and real chemical potential, i.e.~it corresponds to a slice ($N_f=1$ in this case) of the deconfinement critical surface for 
heavy quarks in Fig.~\ref{fig:schem} (right). The two-parameter fit according to tricritical scaling describes
the data perfectly over an astonishingly wide region extending far into real chemical potentials~\cite{OPRW}. 

The idea now is to utilise this scaling behaviour for small quark masses in order to extrapolate to the chiral
limit. The situation is sketched in Fig.~\ref{fig:mc} (right). There is a critical line emanating from the
tricritical point in the Roberge-Weiss plane with known functional behaviour. This line runs towards the
zero mass axis where it must terminate in another, yet unknown tricritical point, which marks the boundary
between a first-order and second-order chiral transition. By mapping out this critical line at finite masses
a controlled extrapolation to the chiral limit will become possible.

\section{Numerical results}

\begin{figure}[t]
\begin{center}
\centerline{
\includegraphics[width=0.55\textwidth]{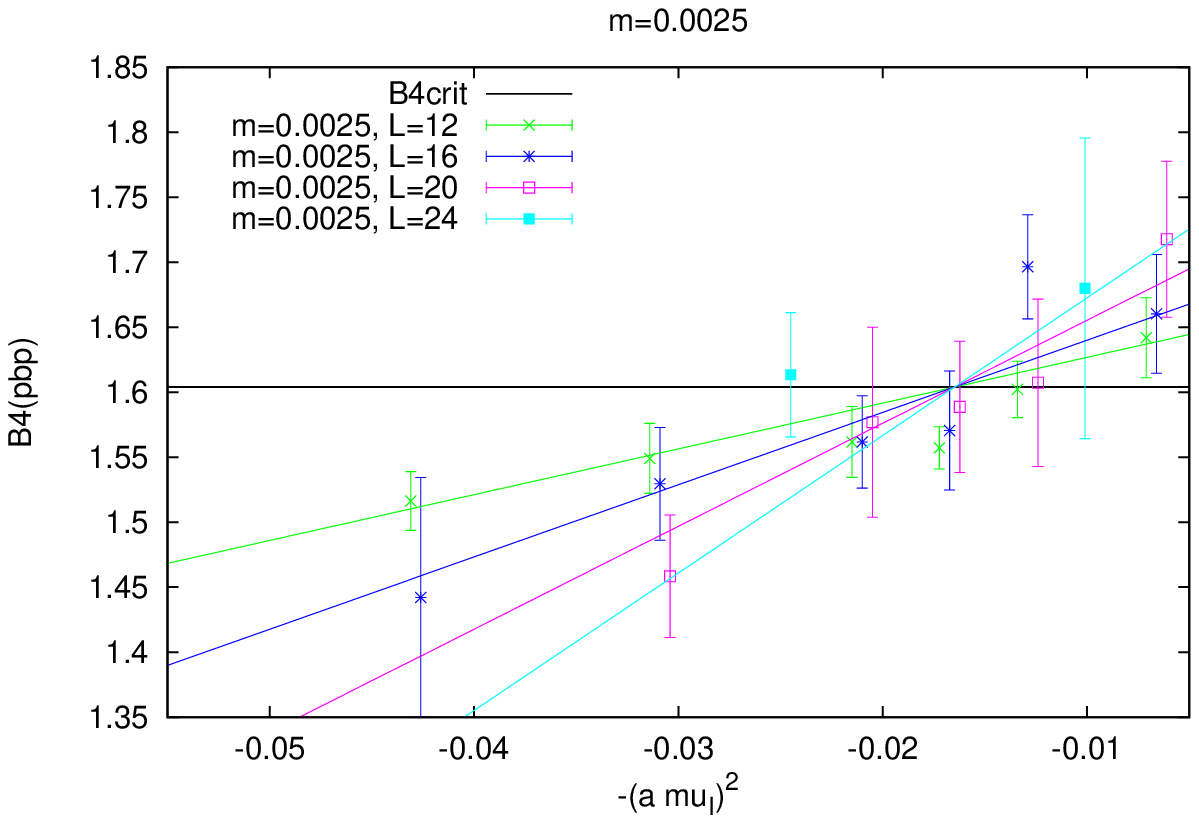}
\hspace*{0.5cm}
\includegraphics[width=0.55\textwidth]{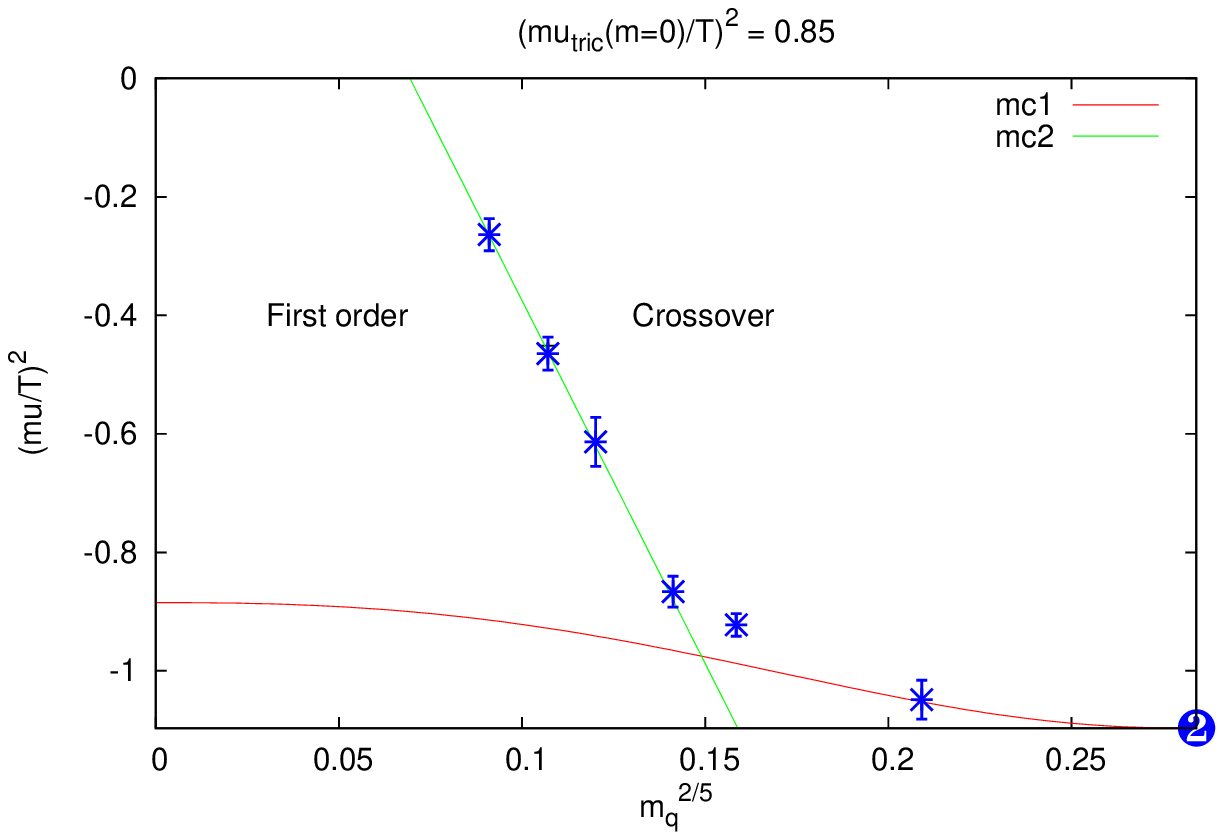}
}
\caption{({\em Left}) Binder cumulant for fixed quark mass as a function of imaginary chemical potential and volume. The intersection marks a critical point. ({\em Right}) Data points are calculated critical points,
the lines are fits to tricritical scaling.
\label{fig:results} }
\end{center}
\vspace*{-0.5cm}
\end{figure}

To map out the critical line in Fig.~\ref{fig:mc} (right), we use the Binder cumulant as an 
observable,
 \be\label{binder}
B_4(m,\mu)=\frac{\langle(\delta X)^4\rangle}
{\langle(\delta X)^2\rangle^2},
\ee
with the fluctuation $\delta X= X-\langle X\rangle$ of the order parameter of interest.
Since we investigate the region of chiral phase transitions, we use the chiral condensate, 
$X = \bar{\psi}\psi$. 
For the evaluation of the Binder cumulant it is implied that
the lattice gauge coupling has been tuned to its pseudo-critical value, $\beta=\beta_c(m,\mu)$,
corresponding to the phase boundary between the two phases.
In the infinite volume limit the Binder cumulant behaves discontinuously, 
assuming the values 1 in a first-order regime,
3 in a crossover regime and the critical value $\approx 1.604$ reflecting the 
3d Ising universality class at a chiral critical point.
On a finite volume the discontinuities are smeared out and 
$B_4(m,\mu)$ passes continuously through the critical value. In its neighbourhood
it can be expanded to leading order,
\be\label{bseries}
B_4(am,a\mu)=
A(am)+B(am) \left((a\mu)^2-(a\mu_c)^2\right) + \ldots,
\ee
with $A(m)\rightarrow 1.604$ for $V\rightarrow \infty$.
An example for our data is shown in Fig.~\ref{fig:results} (left), where we fixed 
the bare mass $am=0.0025$ and scanned in imaginary chemical potential on four different
volumes in order to identify the critical point. All data are fit together fixing the infinite volume value 
for $A$. 

This procedure was carried out for six different values of the quark mass, as shown in 
Fig.~\ref{fig:results} (right). 
We have rescaled the quark mass axis with the appropriate critical exponent in order to display
the scaling and extrapolation more clearly as a straight line.
Four points fall very accurately on the tricritical scaling curve which
can then be used to extrapolate the critical line to the upper tricritical point in the chiral limit, for which
we find the large positive value
\be
\left( \frac{\mu}{T} \right)^2_\text{tric}=0.85(5)\;.
\ee
As the figure illustrates, this implies a definite first-order behaviour for the two-flavour 
chiral phase transition on $N_t=4$ lattices. A crude estimate puts the critical pion mass corresponding
to the second order point at $\mu=0$ to $m_\pi^c\sim$ 60 MeV.
Our result is consistent with the hints of first-order signal seen in earlier scaling studies using staggered fermions on $N_t=4$ \cite{DiG}, as well as with recent 
investigations based on overlap fermions \cite{overlap}.

The presented results have been obtained on coarse $N_t=4$ lattices, corresponding to $a\sim 0.3$ fm.
For $\mu=0$ it is known that the three-flavour chiral first order region in Fig.~\ref{fig:schem} (left) 
shrinks on finer lattices \cite{kim} or with improved actions \cite{ding}. It thus remains to be seen how the chiral critical quark mass identified here evolves with lattice spacing.

\newpage
\section*{Acknowledgements:}
This work is supported by the German BMBF, No.~06FY7100, and the Helmholtz International Center for FAIR within the LOEWE program launched by the State of Hesse.
We thank the Scientific Computing Center at INFN-Pisa, INFN-Genoa,
the HLRS Stuttgart and the LOEWE-CSC at University of Frankfurt
for providing computer resources.

\end{document}